\documentstyle[12pt]{article}
\begin{document}
\begin{center}
{\large\bf Do We See the ``Dark Clouds'' Again? \\}
{\large\bf --- On Some Puzzles in Contemporary Physics}
\end{center}
\vspace{.7cm}
\centerline{Guang-jiong Ni$^*$\footnotetext{$^*$E-mail:
gjni@fudan.ac.cn}}
\vspace{0.2cm}
\centerline{Department of Physics,
Fudan University, Shanghai 200433, P.R. China}
\vspace{0.7cm}
\centerline{Abstract}
Some conceptual problems in physics are discussed. Do we need a change
in the concept of matter structure? Why the ``$i=\sqrt{-1}$'' is 
introduced
to quantum mechanics (QM) essentially? What is the relation between 
QM and
special relativity (SR)? Could we modify the stationary Schr$\ddot{\rm
o}$dinger equation in conformity with SR? How can a particle acquire a
mass? We propose some tentative answer. Their philosophical 
implications
are emphasized.
\vspace*{5mm}

 
\newpage

The cognition of contemporary physics penetrates, on the one side, 
into
the microscopic world at the scale of 10$^{-17}$ cm; on the other 
hand,
it expands to the universe at the scale of 10$^{10}$ light year, i.e.,
10$^{28}$ cm. Meanwhile, it permeates actively into the various fields
of science and technology and play an immense role in pushing them
forward. Now it seems that the pace of progress in physics itself is 
not
as rapid as that in its applications. so the problems is how
could we speculate about the further development of physics in the 
next
century.

In this paper, I would like to join the discussion of this problem 
from
personal point of view.

\noindent 1. Do we see the ``dark clouds'' again?

In 1900, the British physicist Kelven (W. Thomson) had thought that 
the
mansion of classical physics was basically established, the work left
for the physicists in 20th century would be just mending. but he said
again: ``In the remote part of sunny sky of physics, there are two 
small
puzzling dark clouds''. He referred to the Michelson-Morley experiment
and the experiments on black-body radiation. As was well known, from
these two dark clouds emerged the theory of relativity and the quantum
theory. The above story had been cited for many times (e.g. [1]).
However, now I doubt whether it could reflect the overall situation 
at that
time. The physicists (including Kelven), who had good eyesight, should
not fail to pay attention to the three successive discoveries in three
years 1895-1897, the X ray, radioactivity and electron. They actually
raised the curtain of symphony of physics, even the whole science and
technology in the 20th century.

The so called dark clouds imply serious conflict between the new
experimental facts and existing theory. it seems that the 
contradictions
encountered in present day physics is not so clearcut or acute as that
occurred 100 years ago. but at least there are four puzzles which are
generally recognized.

1.1. Why the quarks can not escape from hadrons? (the puzzle of quark
confinement).

1.2. After the discovery of parity violation in 1956, the violation of
CP (i.e. T) invariance was discovered in 1964. But the extent of
violation is only 0.3\%. What does it mean?

1.3 The quasars discovered since the sixties in astronomy are
characterized by their small volumes, high luminosities and strange
phenomena in the change of luminosity. No satisfying  explanation has
been found for them.

1.4. The luminous matter only comprise the minority part of the
universe. the majority part of universe is composed of dark matter. 
What
are they?

I wish to add a theoretical problem generally emphasized:

1.5. What is the essence or origin of mass?

Another theoretical problem may not be so generally established:

1.6. What is the essence of special relativity (SR)? What is the
relation between SR and quantum mechanics (QM)?

In the following, due to my lack of knowledge in 1.3 and 1.4, only the
other four problems will be discussed.

\noindent 2. ``Why an electron can not fall inside?'' and ``Why the
quark can not escape outside?''.

In comparison with the consideration about various possibilities 
beyond
the standard model (SM) in particle physics by some theoretical
physicists, my point of view is rather conservative. I suggest that 
the
first thing should be to digest so profound achievements of the 20th
century physics. In ancient China, Confucius had said that`` to gain 
new
insights through restudying old materials''.

For instance, a fact which forms an acute contrast to problem 1.1
occurred at the beginning of this century. `` Why an electron can not
fall into the nucleus under its attraction?'' It was precisely the 
deep
thinking on puzzle of ``atom stability'' which prompted N. Bohr to
propose his quantum model of atom in 1913. After the establishment of 
QM
by Heisenberg, Schr$\ddot{\rm o}$dinger et al. around 1924, one 
realized
that it is the ``uncertainty relation'' reflecting the ``wave-particle
dualism'' which ensures the stability of atom. When a particle is
compressed, its energy increases. The stronger the attraction force 
is,
the stronger its ``repulsion'' against the compression will be.

Now look at the collision between $e^{+}$ and $e^{-}$ in a collider, 
say
in the BEPC in Beijing. When the energy in the system of center of 
mass
reaches 3.097 GeV, a particle $J/\psi$ may be created.
$J/\psi=(c-\bar{c})$ is considered as a binding state of quark $c$ and
anti-quark $\bar{c}$. However, either $c$ or $\bar{c}$ can not be 
found
in isolated state till now. For example, if the energy is enhanced to
2$\times$1.869 GeV, the bond connecting $c$ and $\bar{c}$ is broken, 
but
a pair of $d$ quark, $d$ and $\bar{d}$ are created immediately. What 
we
see are two separated $D$ mesons, $D^{+}=(c\bar{d})$ and
$D^{-}=(\bar{c}d)$.

So the ``quark confinement'' together with the emergence of
quark-antiquark pair shows a qualitative change in the concept of
structure of matter. In the past, if two particles $a$ and $b$ are
combined into a composite particle A with binding energy B being much
smaller than the rest energy E$_{0}$, we can say that a and b are
constituent particles or ``building blocks'' in A=(ab). This is 
because
the difference between the binding state of a or b in A and the free
state after separation is small( $B/E_{0} \ll 1$) . One could often
neglect the difference.
Now the situation is no longer true. The fact of ``quark confinement''
tells us that the structure of matter is essentially not piled up by
particles but a structure of something else [2].

It is intimately related to problem 1.5. The rest mass of $e^{+}$ or
$e^{-}$ is only 0.511 Mev, less than 1.65$\times 10^{-4}$ of $J/\psi$
mass. So $J/\psi$ is not hiding inside the $e^{-}$ or $e^{+}$, but is
excited out of vacuum during the collision. But what object is 
excited?
In our opinion, it is nothing but the ``fundamental contradictions'' 
in
the nature. The important thing is: a ``contradiction'' is massless 
and
invisible before it is excited. This was a conjecture by G.W.F. Hegel
and F. Engels, now is actually verified by experiments in physics [3].

It seems to us that only the concept of ``contradiction'' is capable 
of
grasping, in 100 years time span, two features of matter structure ---
``Why an electron can not fall into nucleus?'' and ``why the quark can
not escape from a hadron?'' they seems opposite but virtually are
complementary to each other. As this kind of concept is beyond the 
scope
of usual atomism familiar to western physics community, physicists had
underestimated the ``repulsiveness'' of contradiction again and again,
now are still underestimating the ``identity'' of it.

\noindent 3. The theory of ``primary gas'' and new ``ether''.

For further exploiting the above point of view , let us return back to
QM. The wave function of a free particle with momentum $\vec{p}$ and
energy $E$ is described by
\begin{equation}
\psi \sim exp[\frac{i}{\hbar}(\vec{p}\cdot\vec{x}-Et)]
\end{equation}
In our explanation, being an abstract representation of contradiction
between particle and its environment [1], $\psi$ consists of two 
parts:
\begin{equation}
\psi=Re\psi+iIm\psi
\end{equation}
Under a phase transformation, it reads:
\begin{equation}
\psi\rightarrow \psi^{'}=e^{i\theta}\psi
\end{equation}

\begin{eqnarray}
Re\psi &\rightarrow& Re\psi^{'}=\cos\theta Re\psi-\sin\theta Im\psi
\nonumber \\
Im\psi &\rightarrow& Im\psi^{'}=\sin\theta Re\psi+\cos\theta Im\psi
\end{eqnarray}
Note that: (a) Both Re$\psi$ and Im$\psi$ are real, but they are all
nonobservables. (b) The distinction between them is necessary. One 
side
regards the existence of other side as the premise of existence of
itself. They are indivisible. Any one of two sides can not exist
individually. (c) They can be transformed each other as shown in 
Eq.(4)
while keeping $|\psi|^{2}$ invariant.

In Ref.[4] (earlier by Schwinger et al. [5]), it was pointed out that
for an antiparticle with momentum $\vec{p}$ and energy $E(>0)$, the 
wave
function should be written as:
\begin{equation}
\psi_{c}\sim \exp[-\frac{i}{\hbar}(\vec{p}\cdot\vec{x}-Et)]
\end{equation}
Accordingly, a particle is not absolutely pure. The ingredient of
``antiparticle state'' hiding inside will increase with the velocity 
of
particle. then the mass-energy relation $E=mc^{2}$ can be derived.

We would like to add that mass is a low energy phenomenon familiar in
daily life. So its origin, i.e. the essence of $E=mc^{2}$, must be
stemming from one (not two) universal, simple but subtle law, which in
turn must
be already contained in existing experimental and theoretical 
knowledge.
As a contrast, when Einstein established the SR, he even did not 
mention
the Michelson-Morley experiment explicitly. When Einstein further
established the general relativity, besides the SR and the 
gravitational
law, he was mainly depending on the experimental fact that `` all 
freely
falling body on the earth have the same acceleration $g\approx 9.8
m/sec^{2}$'' which was seen by every one.

The postulate implied in Eqs. (1) and (5) is really very simple and
universal --- ``space-time inversion ($\vec{x}\rightarrow -\vec{x}$,
$t\rightarrow -t$) is equivalent to particle-antiparticle
transformation''. Basing on this fundamental symmetry, we derived a
``relativistic stationary Schr$\ddot{\rm o}$dinger equation for many
particle system'' [6] with eigenvalue $\epsilon$ related to the 
binding
energy B of system as
\begin{equation}
B=Mc^{2}[1-(1+\frac{2\epsilon}{Mc^{2}})^{1/2}]
\end{equation}
($M=\sum_{i=1}^{n} m_{i}$), showing the consistency between SR and QM 
in
essence. It is well known in QM that the operators for particle read
(see Eq. (1))
\begin{equation}
\vec{p}\rightarrow -ih\nabla, E\rightarrow ih\frac{\partial}{\partial 
t}
\end{equation}
Now Eq.(5) implies that for an antiparticle:
\begin{equation}
\vec{p}_{c}\rightarrow ih\nabla, E_{c}\rightarrow
-ih\frac{\partial}{\partial t}
\end{equation}
According to the theory of ``primary gas'' (Yuan-qi) in Chinese
philosophy [7], Re$\psi$ and Im$\psi$ could be named as ``yin''
(feminine or negative) and ``yang'' (masculine or positive). Eqs. (1)
and (5) show that two opposite coupling modes of them correspond to
particle and antiparticle. In 1905, the old theory of ``ether'' was
abandoned for establishing SR. It seems to us that the revival of new
theory of ``ether'' basing on ``Yuan-qi'' is inevitable. Now it is the
time to combine two kinds of wisdom of both eastern and western 
philosophy.

\noindent 4. Is there an ``arrow of time''?

In defining the so called time reversion (T) in QM, one not only sets
$t\rightarrow -t$, but also takes a complex conjugation:
$\psi\rightarrow\psi^{*}$, for ensuring the invariance of 
Schr$\ddot{\rm
o}$dinger equation. Hence, actually, it implies an equivalence:[8]
\begin{equation}
\psi(\vec{x},t)\sim\psi^{*}(\vec{x},-t)
\end{equation}
Obviously, the usual stationary(eigen) state does satisfy Eq.(9).  The
discovery of CP (i.e. T) violation in neutral K system since 1964 has 
a
possible explanation within the framework of SM. If the eigenstate of
quark in strong interaction not coincides with that in weak 
interaction,
they are linking together via an unitary transformation, the CKM 
matrix.
Then a phase angle in the matrix may account for the 0.3\% discrepancy
in CP violation. If this explanation can be further verified in
experiments, then the nonconservation of T reversal does mean a 
peculiar
problem in particle physics (there are two kinds of eigenstates for
quarks, there exists flavor mixing in the weak interaction 
eigenstates),
but has nothing to do with the basic symmetry in space-time.

C. N. Yang and C. P. Yang had proved that the violation of T reversal
has nothing to do with the macroscopic nonreversibility inferred by 
the
second law in thermodynamics [9]. In our point of view, with respect 
to
the problem of time reversal, the so called ``Loschmidt paradox''
between ``microscopic reversibility and macroscopic 
nonreversibility'' does
not exist at all. This is because the symmetry discussed in previous
section implies that the time reversal must be accompanied by the
transformation of matter to antimatter. In some sense, the ``arrow of
time'' already exists at microscopic level. So during the transition
from microscopic scale to macroscopic one, instead of the problem that
``how can an arrow of time emerge from none'', we should ask ``how can
it display explicitly from implicit existence''. We think that the
entropy equals zero when a macroscopic system is in a quantum coherent
state like that in superconductivity or superfluidity. Once the 
quantum
coherence is destroyed, the entropy increases and the arrow of time
emerges explicitly [10].

\noindent 5. The unity of opposites between individual and its
environment.

We set $E=m_{0}c^{2}-i\hbar/2\tau$ in Eq.(1), yielding
\begin{equation}
\psi(t)\sim \exp [-\frac{i}{\hbar}m_{0}c^{2}t-\frac{t}{2\tau}],
|\psi(t)|^{2}\sim \exp(-\frac{t}{\tau})
\end{equation}
with $\tau$ being the lifetime of particle. Being the ``imaginary part
of mass'', the decay constant $\frac{1}{\tau}=\Gamma$ is determined 
by its
environment. Different nuclei have different life time in Beta decay.
Correspondingly, the ``real part of mass'', $m_{0}$, should also 
depend
on its environment. After studying the quantum field theory (QFT) for 
40
years, I began to realize that I was totally wrong to think that the
observed mass is generated from an intrinsic ``bare mass''. Now I
understand that the calculation of ``self-energy'' in perturbative QFT
has nothing to do with the mass generation. Can we create a mass from 
a
massless model like NJL model [11]? Yes, the outcome turns out to be
either no mass scale or two mass scales, but never one mass scale. The
reason lies in the fact that besides the mass of particle, another
mass scale is needed as a standard weight, which is provided by the
phase transition of vacuum( {\it e.g.} the extra constant $\Lambda
\approx 200 MeV$ in QCD, being a necessary complement to the 
Lagrangian,
plays the role as a standard weight). [12,13]

We all understand that ``Many body system is comprised of individuals.
If there is no individual, there is no system''. But the above 
statement
merely forms the half of truth. The another half is more important:
``The existence of individual is ensured by its environment. If there 
is
no certain environment, there is no certain individual''. Chinese
philosophy talked about ``oneness of heaven and man''. It makes sense.

\noindent 6. Infinity is essentially different from finiteness

The pertubation theory in QFT is expanded with respect to the loop
number L in Feynman diagram. The tree level corresponds to L=0. Once 
the
quantum radiation correction is taken into account in L=1, 2, ...
calculations, one encounters the divergence, i.e. $\infty$, 
immediately.
For dealing with it, one devises many tricks like the counter term and
bare parameter, etc. I don't believe in these kind of trick any 
longer.
After learning the relevant literatures and beginning from a former
graduate student, J-f Yang, we now adopt a new simple trick. Then
instead of divergence, we have some arbitrary constants C$_{i}$. 
Neither
counter term nor bare parameter is needed. The renormalization is 
simply to
fix these C$_{i}$ by experiments. So there is also no arbitrary 
running
mass scale $(\mu)$ after renormalization, [14-17].

Therefore we begin to realize that the appearance of $\infty$ is no 
more
than a signal. It is essentially a warning: We expected too much, it 
is
impossible to evaluate the mass or charge in pertubative QFT. Only 
after
we confess that our ability is limited, can we make the target of our
theory more clearcut and the latter becomes more predictive. For
example, recently we calculated the Higgs mass in SM to be 138 GeV 
[18],
which is in conformity with the outcome of phenomenalogical analysis 
on
present experiments. [19]

Moreover, ``$\infty$'' is by no means a very large fixed number. If
performing the perturbation calculation in QFT to certain order of L, 
no
matter how large it is, we still have perturbative theory, which is 
not
quite different from the tree level calculation in essence. In
particular, the mass of a free particle remains the same. Only after
taking L$\rightarrow\infty$, i.e. performing the nonperturbative
evaluation, can we try to discuss the problems like mass generation.
Then, as mentioned in previous section, the phase transition of vacuum
will occur. The number of degrees of freedom N in the environment of a
particle is not a large number, but N$\rightarrow\infty$. An 
elementary
example is the geometric series: $S_{n}=1+r+r^{2}+\cdots+r^{n}$ is
essentially different from $\lim_{n\rightarrow\infty}S_{n}=(1-r)^{-
1}$.

We believe that $L\rightarrow\infty$, $N\rightarrow\infty$, this is 
the
difficult problem in theoretical physics challenging us. We are always
facing the dilemma that ``the world is infinite while our knowledge
remains finite''. Hence it is inevitable that every theoretical model 
is
limited to its boundary of applicability, where some singularity is
destined to emerge, e.g., the black hole in general relativity.
If a theory really gets rid of any singularity, it must be trivial in
the sense of being meaningless or even wrong, just like what said by 
the famous
Liouville theorem in the complex variable function theory.

\noindent 7. ``We are actors, and spectators as well''

Finally, I don't think that the basic research of physics in the next
century will enjoy more brilliant achievement than that gained in the
20th century. The reason is plain: the majority of basic laws 
governing
the existing matter on the earth seems comparatively clear. However, 
the
applications of physics, the combination of physics with other science
and technology, especially astronomy and life science, are just in
blooming. During this process we will discover the harmony form large
cosmos to tiny world. We will face the infinite world consciously and
cognize ourselves more consciously. As Weisskopf said [20], we, who 
are
living in the 20th century, are privileged to witness the most 
exciting
phase in the evolution process of living beings. It is on the earth 
the
greatest adventure of the universe takes place --- that nature in the
form of man begins to understand itself. Once upon a time, Bohr 
said:``
In searching for the harmony of life, one should never forget that in
the drama of existence, we ourselves are actors, and spectators as
well''. While there are so many challenges facing our mankind, let us
unite and work together for a better world in the next century.

\newpage

\centerline{REFERENCES}

\noindent [1] Guang-jiong Ni and Hong-fang Li, Modern Physics, 
Shanghai
Scientific \& Technical Publisher, (1979).

\noindent [2] Guang-jiong Ni, {\it What is the structure of elementary
particles}? Physics, 8(4), 378 $\sim$ 380 (1979).

\noindent [3] G-j Ni, {\it Philosophical
thinking on physics}, Fudan Journal (Social Sciences Edition), (2),
3 $\sim$ 5 (1997).

\noindent [4] G-j Ni, {\it Relation between space-time inversion and
particle-antiparticle transformation}, Fudan Journal (Natural Science 
Edition),
(3-4), 125 $\sim$ 134 (1974); G-j Ni and S-q Chen, {\it On the 
essence of
special relativity}, ibid, 35(3), 325 $\sim$ 334 (1996); G-j Ni and S-
q
Chen, preprint, Internet Hep-th/9508069, (1995).

\noindent [5] J. Schwinger, Proc. nat. Acad. Sc. U.S., 44, 223 (1958);
E.J. Konopinski and H. M. Mahmaud, Phys. Rev. 92 , 1045 (1953).

\noindent [6] G-j Ni and S-q Chen, {\it Relativistic stationary
Schr$\ddot{\rm o}$dinger equation for many-particle system}, Fudan
Journal (Natural Science Edition), 36 (1997), (3)247-252; G-j Ni,
preprint, Internet, hep-th/9708156.

\noindent [7] $<<$Ci-Hai$>>$ (Sea of Words), Shanghai Dictionary 
Publisher
(1979), see ``Yuan-qi'' and ``Tai-ji''; Z-s He, Science in China, (5)
445 $\sim$ 455 (1975).

\noindent [8] J. J. Sakurai, Modern Quantum mechanics, the Benjamin
Cummings Publishing Company, Inc., sec. 4.4, (1985).

\noindent [9] C. N. Yang and C. P. Yang, Transaction of the New York
Academy of Science, 40, 267 (1980).

\noindent [10] G-j Ni, S-q Chen and G-s Zhou, Acta Phys. Sinica, 31, 
585
(1982).

\noindent [11] Y. Nambo and G. Jona-Lasinio, Phys. Rev. 122, 345 
(1961).

\noindent [12] G-j Ni, J-f Yang, D-h Xu and S-q Chen, Commun. Theor.
phys. 21, 73 (1994).
   
\noindent [13] G-j Ni and S-q Chen, $<<$ Levinson Theorem, Anomaly 
and the
Phase Tansition of Vacuum$>>$, Shanghai Scientific \& Technical
Publishers, Chap.8.2, (1995).

\noindent [14] Ji-feng Yang, Thesis for PhD, Fudan Univ.,
unpublished, (1994). Preprint, hep-th/9708104.

\noindent [15] J-f Yang and G-j Ni, Acta Phys. Sinica, 4, 88 (1995).
G-j Ni and J-f Yang, Phys. Lett. {\bf B} 393(1997), 79.

\noindent [16] G-j Ni and S-q Chen, accepted by Acta Physica Sinica
(overseas version), Internet hep-th/9708155.

\noindent [17] G-j Ni and Haibin Wang, preprint, Internet hep-
ph/9708457.

\noindent [18] G-j Ni, S-y lou, W-f Lu and J-f Yang, preprint, 
Internet
hep-ph/9801264.

\noindent [19] J. Gillies, Warsaw Conference, CERN Courier, 36(7),
1 $\sim$ 8 (1996).

\noindent [20] V. F. Weisskopf, $<<$ Physics in the Twentieth 
Centurys:
Selected Essays$>>$, The MIT Press, (1972).

\noindent [21] This work was supported in part by the NSF in China.

\noindent [22] This paper is a revised English version of its
Chinese counterpart, which is published in the Journal ``Physics
Bulletin'', (1997), No. 5, 39-42.
\end{document}